\documentclass[conference]{IEEEtran}
\renewcommand\IEEEkeywordsname{Keywords}

\usepackage[pdftex]{graphicx}
\usepackage{amsmath}
\usepackage{amssymb}
\usepackage{caption}
\usepackage[font=small]{caption}
\newcommand{\rpm}{\raisebox{.2ex}{$\scriptstyle\pm$}}
\AtBeginDocument{%                     % this will reduce spaces between parts (above and below) of texts within a (sub)section to 0pt, for example - like between an 'eqnarray' and text
  \setlength\abovedisplayskip{0pt}
  \setlength\belowdisplayskip{0pt}}
  
\begin{document}
\captionsetup[figure]{name={Figure},labelsep=period}

\title{\LARGE \bf Multi-Contact Force-Sensing Guitar for Training and Therapy}
\author{\authorblockN{Zhiyi Ren\authorrefmark{1}, Chun-Cheng Hsu\authorrefmark{1}, Can Kocabalkanli\authorrefmark{1}, Khanh Nguyen\authorrefmark{1}, Iulian I. Iordachita\authorrefmark{1},\\Serap Bastepe-Gray\authorrefmark{2}\authorrefmark{3}, Nathan Scott\authorrefmark{1}}
\authorblockA{\authorrefmark{1}Department of Mechanical Engineering, Johns Hopkins University, Baltimore, MD, USA}
\authorblockA{\authorrefmark{2}The Peabody Institute, Johns Hopkins University, Baltimore, MD, USA}
\authorrefmark{3}Department of Neurology, School of Medicine, Johns Hopkins University, Baltimore, MD, USA\\Email: zren5, chsu40, ckocaba1, knguye44, iordachita@jhu.edu, sbastep2@jhmi.edu, nscott@jhu.edu}

\maketitle

% ================
% # Abstract     #
% ================
\begin{abstract}
Hand injuries from repetitive high-strain and physical overload can hamper or even end a musician's career. To help musicians develop safer playing habits, we developed a multiple-contact force-sensing array that can substitute as a guitar fretboard. The system consists of 72 individual force sensing modules, each containing a flexure and a photointerrupter that measures the corresponding deflection when forces are applied. The system is capable of measuring forces between 0-25 N applied anywhere within the first 12 frets at a rate of 20 Hz with an average accuracy of \rpm 0.4 N and a resolution of 0.1 N.  Accompanied with a GUI, the resulting prototype was received positively as a useful tool for learning and injury prevention by novice and expert musicians.
\end{abstract}

% We also developed a GUI to visualize and display the measured forces in real time and to save and process said measurements.

\begin{IEEEkeywords}
Force sensing, photointerrupter, sensing array, flexure deflection, injury prevention
\end{IEEEkeywords}

% ========================
% # I. Introduction      #
% ========================
\vspace{2mm}
\section{Introduction}
%\vspace{3mm}

%Repetitively overloading the fingers in contorted positions inevitably leads to injuries among guitarists. For guitarists, harmful playing habits are made worse when fingers are repetitively overloaded in contorted positions. Many professional musicians never fully recover to their highest level after getting injured.%
% For guitarists, repetitively overloading the finger joints in contorted positions %

% For guitarists, the problem lies in repetitively overloading 

Many musicians never fully recover to play at their highest level after suffering from injuries. For guitarists, injuries are common even in experts due to the repetitive overloading of finger joints developed through years of harmful playing habits \cite{winspur1997musician,rigg2003playing}. By measuring the dynamic forces applied by the players on the frets and strings of the guitar, we aim (1) to warn musicians in real time during practice when their grip is too strong, (2) and to enforce correct force patterns.

In a preliminary experiment involving one of the senior authors, two 6-axis ATI Nano 17 force/torque sensors were mounted on a guitar. Despite valuable insight, the system modified the original shape of the guitar significantly and only measures the gross force on the fretboard, thus unable to distinguish individual finger forces. Similar work in force sensing on musical instruments suffers the same issues \cite{hori20133, kinoshita2009left}. In our work, we address both shortcomings by developing a modular sensing array that can be packed into a customized, 4mm thick fretboard (Fig. 1). It can measure forces at 72 intersections of frets and strings, and can be mounted onto a classical guitar easily with little change in appearance.

Typical sensing elements used in force sensing arrays include piezoresistive sensors \cite{papakostas2002large}, capacitive sensors \cite{hoshi2005sensitive}, and resistive composites like Velostat \cite{giovanelli2016force}. Both piezoresistive sensors and Velostat are known for their inherent drift and hysteresis. Capacitive sensors require complicated circuitry and customization processes \cite{cutkosky2008force, wu2018capacitive}. Instead, we adopted optical proximity sensors as the sensing elements, inspired by \cite{goldberg2015force}. The sensors are low-cost and lightweight, and require minimal space for operation. The nature of the technology also eliminates the possibility of drift or hysteresis. In our work, an array of photointerrupters (Sharp GP2S60) measures the distance to flexures that deflect proportionally to the force applied. While light isolation is necessary to mitigate interference from adjacent modules, specific design considerations detailed in Section II were developed to resolve this issue. Though 3-axis force measurement would be ideal and beneficial for our purposes (to measure forces from vibratos, for instance), we focus only on the vertical component of the force in this preliminary work. The preliminary experiment using ATI Nano 17 sensors reveals that a sensing range of 0-25N is sufficient. After calibrating and validating to the full sensing range, we achieved an average error of \textless0.4N RMSE (root mean squared error) and a worst error of \textless5\% FSO (full scale output) at a resolution of 0.1N on most sensing modules.

\begin{figure}[h]
\centering
\captionsetup{justification=centering}
\includegraphics[width=\columnwidth]{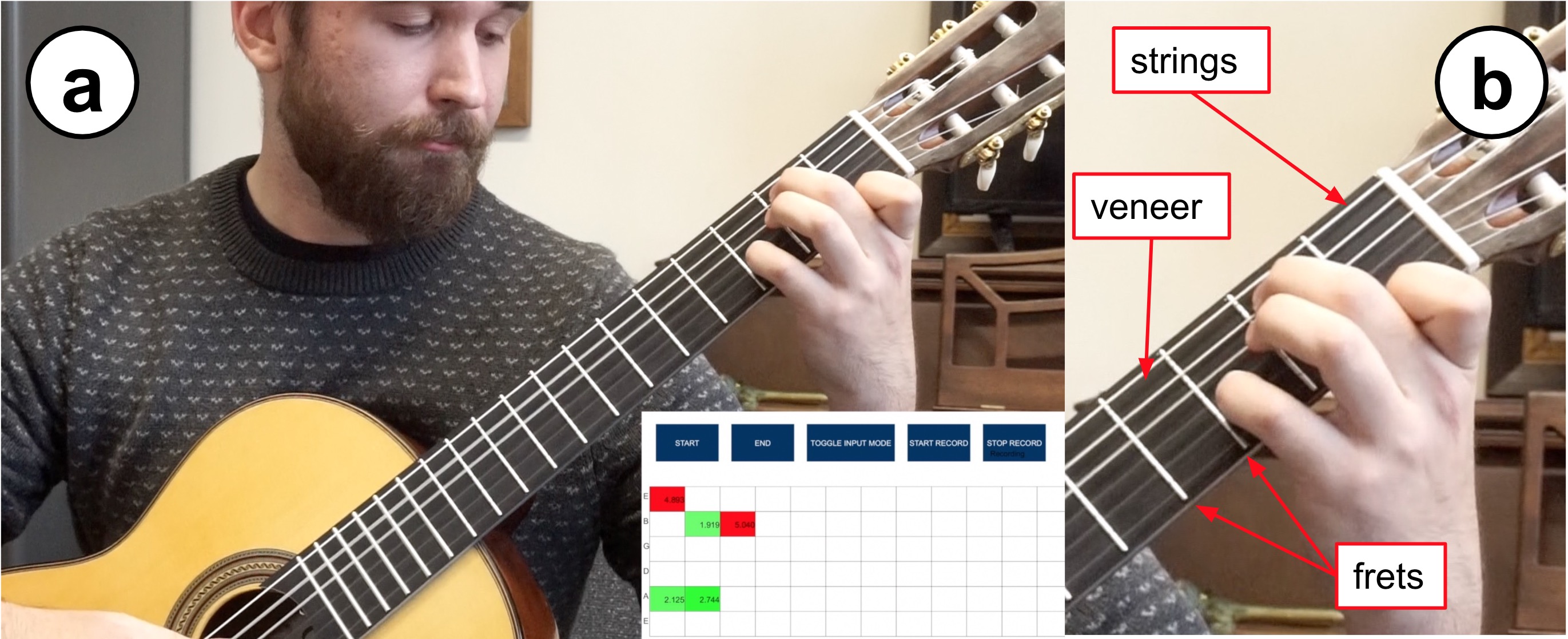}
\caption{(a) GUI showing forces at each string-fret intersection during user testing. Tiles turn red when user presses above a specified force threshold. (b) Fretboard terminology. Also note the contortion of the hand.}
\label{user}
\end{figure}

% The next section demonstrates the implementations of the sensing fretboard in details. Section III documents the testing setup that we built to obtain the calibration and validation results. Section IV finishes the paper with a conclusion, brief summary of the user testing, and future directions.

% ====================
% # II. Implementation #
% ====================
%\vspace{2mm}
\section{Implementation}
%\vspace{3mm}
\subsection{Sensing module characteristics}
The system is comprised of 72 sensing modules, each containing a pair of flexure and photointerrupter. Each photointerrupter has an emitter and a detector: infrared light is emitted by the emitter, reflected off from a fixed surface, and received by the detector. The output is proportional to the amount of light received, which corresponds to the distance to the surface when the environment is dark. As the force is applied onto the flexure, the distance from the flexure to photointerrrupter changes.
\subsection{Flexure design}
The purpose of the flexures is twofold: to provide a measurable deflection and a fixed boundary condition for which the string can vibrate on to create the standing wave we hear as musical notes. Through our finite element analysis in ANSYS \cite{ansys}, we designed the thickness of the flexure such that it elastically deflects 0.2mm under 25N of force. To create the fixed boundary condition, the flexures are designed with the exact same crown profile as a regular fret, with the additional ability to independently deflect under each of the 6 strings (Fig. 2). A compromise has been made to replace the bottom 7 frets (less often used) with non-functional ones in order to house electronic components.
\begin{figure}[h]
\centering
\captionsetup{justification=centering}
\includegraphics[width=0.8\columnwidth]{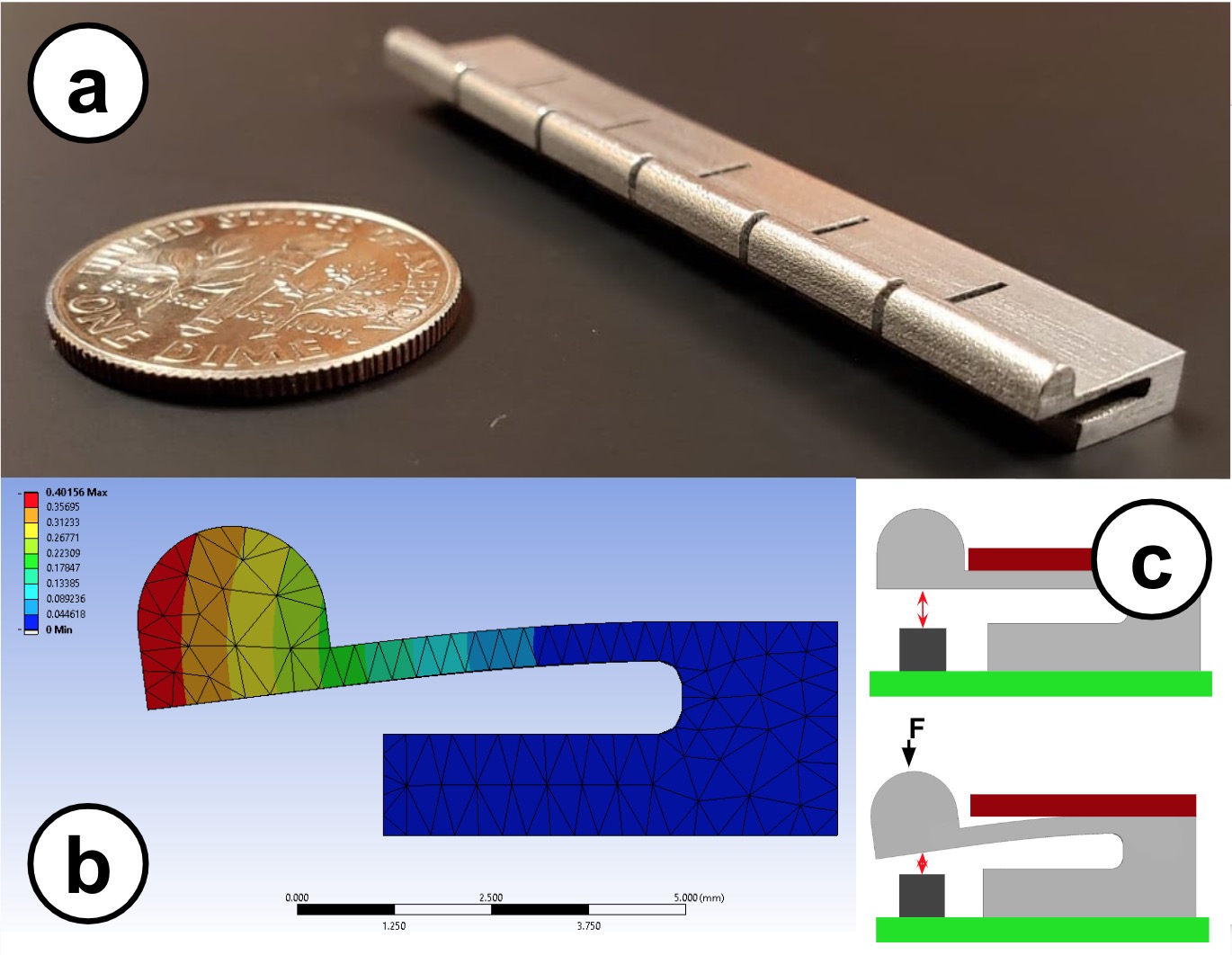}
\caption{(a) A single flexure unit with the profile of a regular fret. (b) FEA analysis under static load. (c) Deflection process when force is applied.}
\label{circuits}
\end{figure}
\subsection{Electronics design}
All electronics components are mounted on a single piece of PCB spanning the whole fretboard. A chain of shift registers (Texas Instruments SN74HC164) activate each of the 72 photointerrupters in sequence. The output signal of each photointerrupter, passing through a unity gain amplifier, is then fed into a differential amplifier (Texas Instruments LM324). There are 6 output lines corresponding to the 6 strings (Fig 3). A digital-to-analog converter (Texas Instruments TLV5638) provides the reference voltage for each differential amplifier. The whole system is controlled by a microprocessor (Texas Instruments MSP430F5342), which has an internal, 12-bit analog-to-digital converter that reads the sensor signal.

The system is powered through a USB connection through a Micro-USB port installed at the bottom of the guitar. By connecting the fretboard to a computer, the user can visualize real-time force measurements at all locations at a rate of 20Hz and save them in text files with a custom GUI made in Processing language \cite{processing}.

\begin{figure}[h]
\centering
\captionsetup{justification=centering}
\includegraphics[width=0.9\columnwidth]{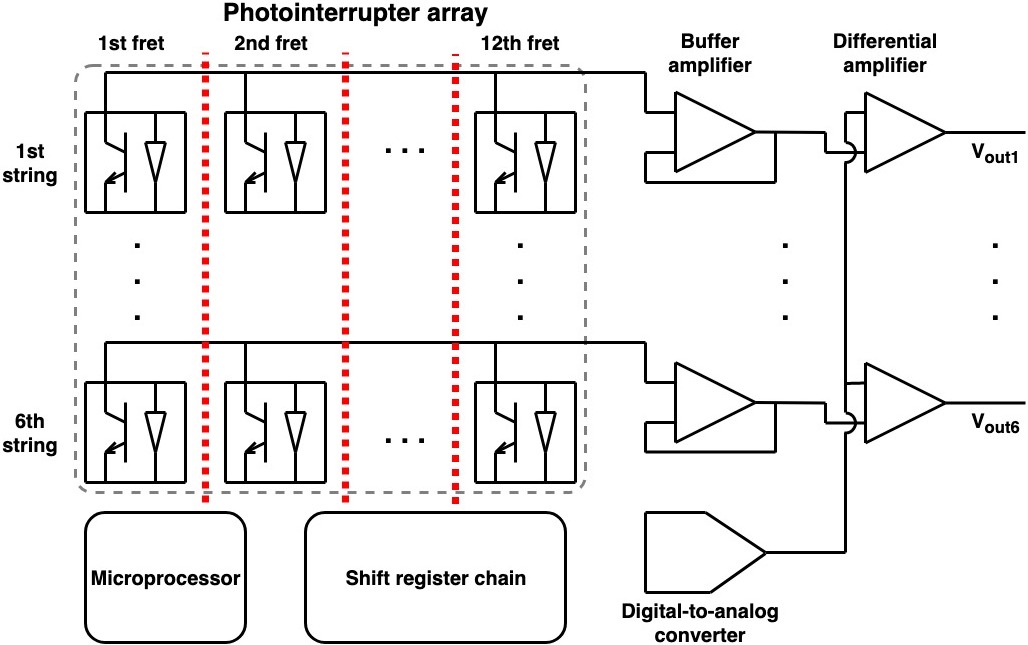}
\caption{Simplified diagram of the circuits. The red dotted lines in the photointerrupter array indicate light isolation between frets.}
\label{circuits}
\end{figure}

\begin{figure}[h]
\centering
\captionsetup{justification=centering}
\includegraphics[width=0.9\columnwidth]{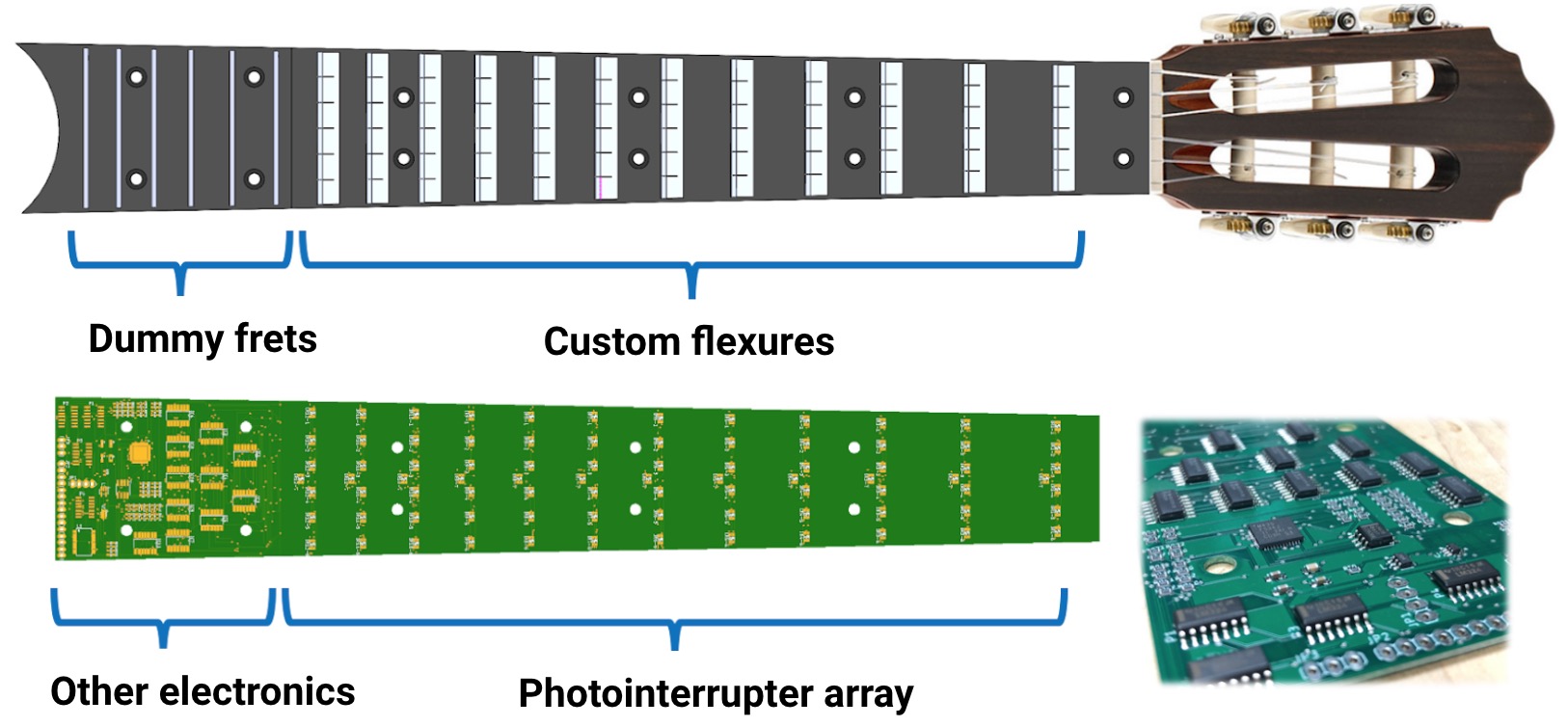}
\caption{Top view of the fretboard and the PCB underneath. The top 12 frets measure forces with photointerrupters. The lower 7 dummy frets house the rest of the electronics.}
\label{pcb}
\end{figure}

\subsection{Assembly}
The aluminum flexures are glued onto the PCB surface with Araldite glue (Huntsman, The Woodlands, TX) such that each photointerrupter sits directly under each flexure. The PCB fits into an aluminum case that houses the flexures and covers the remaining area of the PCB. The case is screwed onto the guitar neck with inserts, and a wooden veneer is glued onto the outermost surface.

\begin{figure}[h]
\centering
\captionsetup{justification=centering}
\includegraphics[width=\columnwidth]{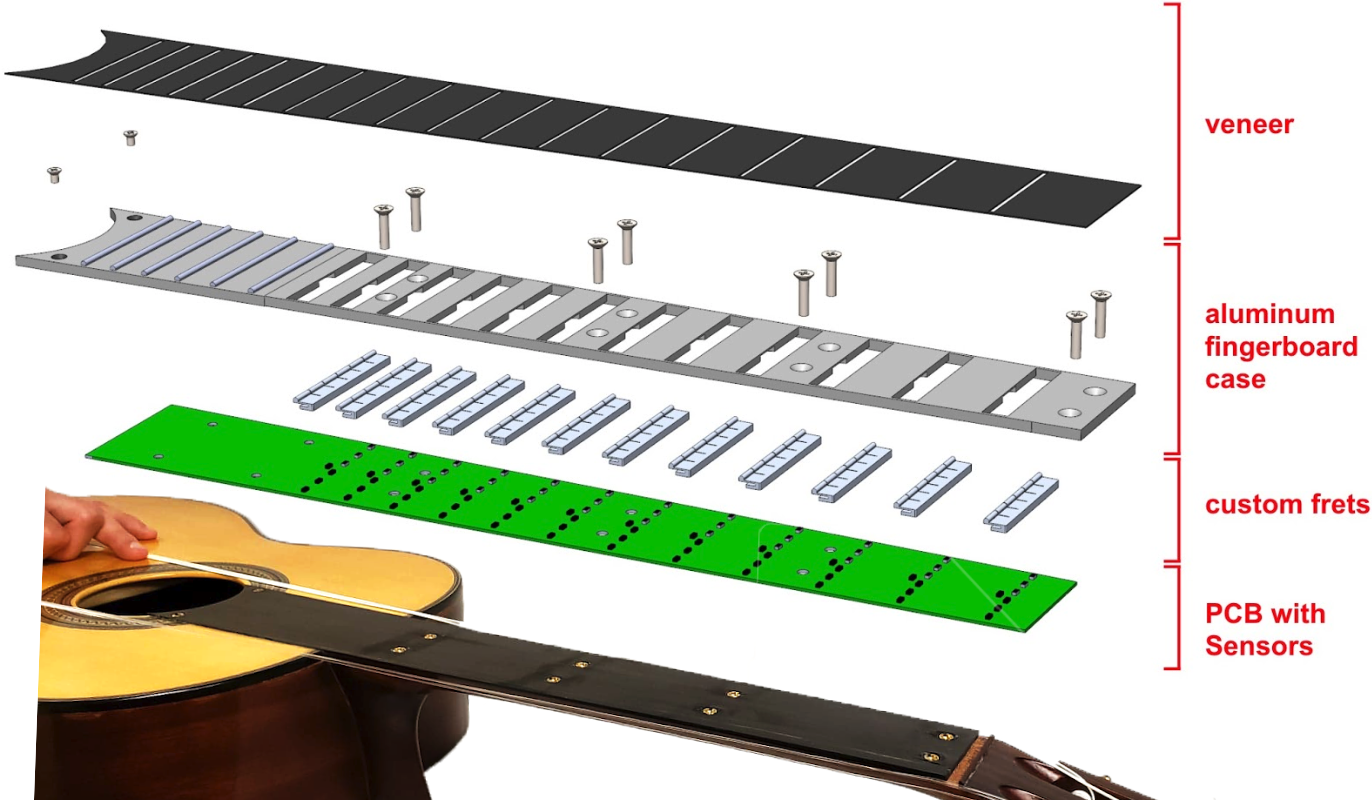}
\caption{Exploded views of the sensing fretboard. The whole system is contained in a custom, 4mm thick fretboard, which can be mounted on a classical guitar with screws.}
\label{exploded}
\end{figure}

\subsection{Light isolation}
The mechanical and electronic designs ensure that there is no light interference between photointerrupters, which is critical for the performance of a optical sensing array. The aluminum case fills the space between the frets thus no light can transmit among them, as shown with the red dotted lines in Fig. 3. This allows the 12 photointerrupters on the same string to share a single output bus. Since each of the 6 photointerrupters within the same fret are on separate output lines and are activated in sequence, there is no light interference within the fret (among the 6 adjacent sensors) either. In addition, the veneer on top of the case also protects the photointerrupters from any external light source.

% ==========================
% # III. Experiments and Results #
% ==========================
\vspace{2mm}
\input\section{Experiments and Results}
%\vspace{3mm}
\subsection{Testing setup}
We developed a testing setup that can move freely along both the string and fret directions to calibrate and validate each of the 72 sensing modules efficiently. By precisely displacing a micron stage we apply varying forces onto the the crown on the sensing unit (Fig. 2c). A Sparkfun TAL220B load cell (\rpm 0.025 N accuracy), calibrated with known weights, is mounted at the tip of the micron stage to provide the ground truth reading against the output of the sensing module. 
\begin{figure}[h]
\centering
\captionsetup{justification=centering}
\includegraphics[width=\columnwidth]{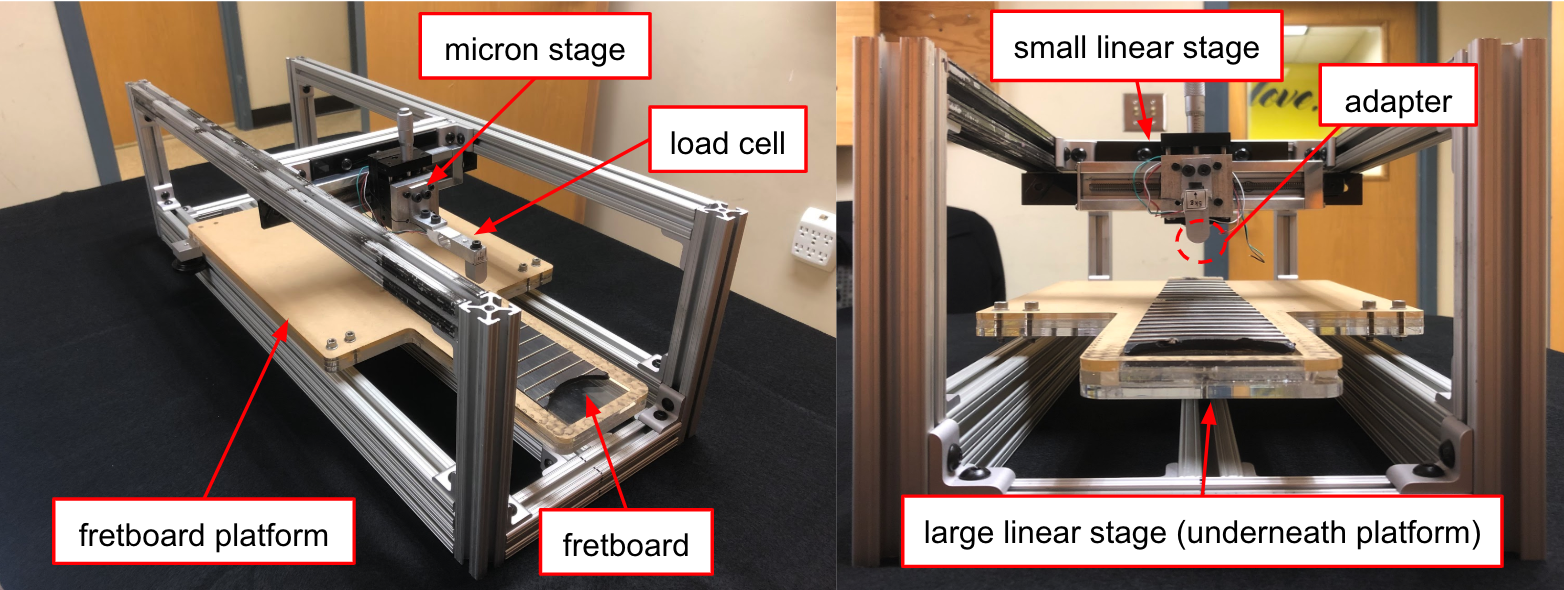}
\caption{Testing rig used to apply force across all 72 modules.}
\label{rig}
\end{figure}

\begin{figure}[h]
\centering
\captionsetup{justification=centering}
\includegraphics[width=0.8\columnwidth]{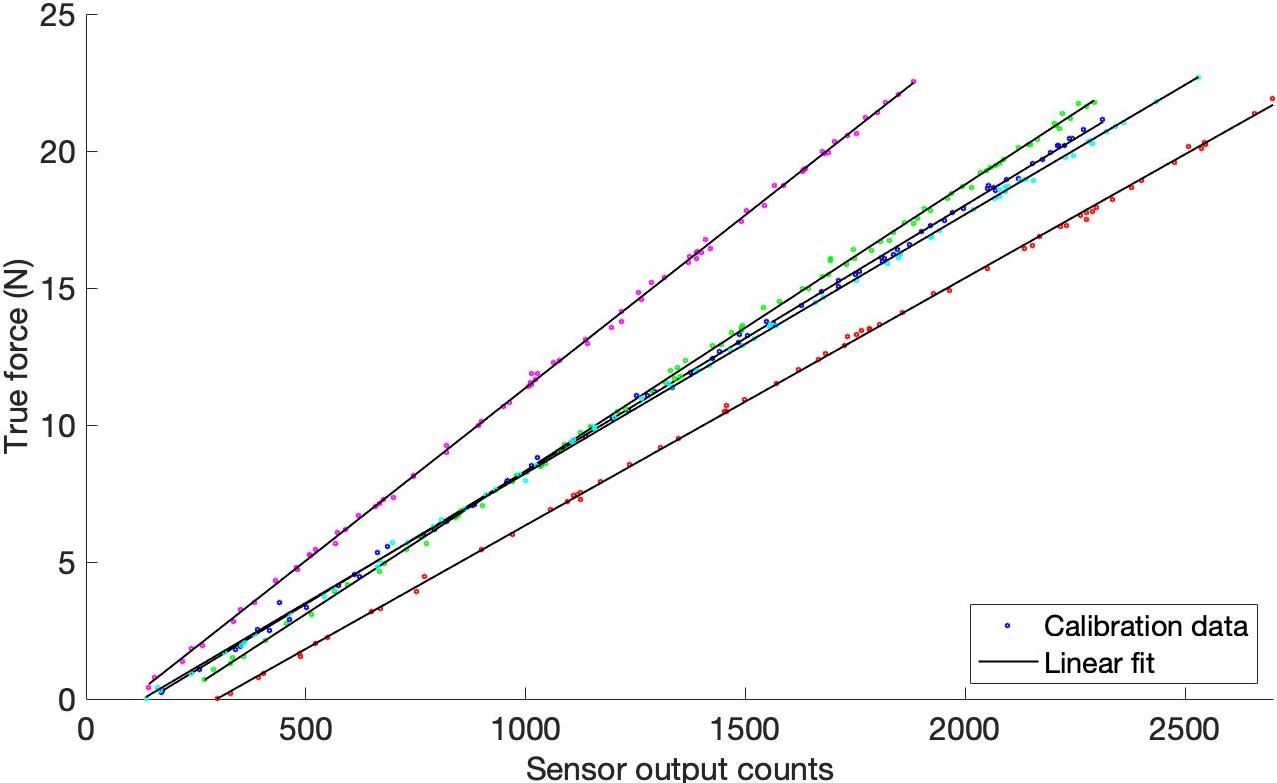}
\caption{Calibration results of five randomly chosen modules. The coefficient of determination for each is higher than 0.99.}
\label{calibration}
\end{figure}

\begin{figure}[h]
\centering
\captionsetup{justification=centering}
\includegraphics[width=\columnwidth]{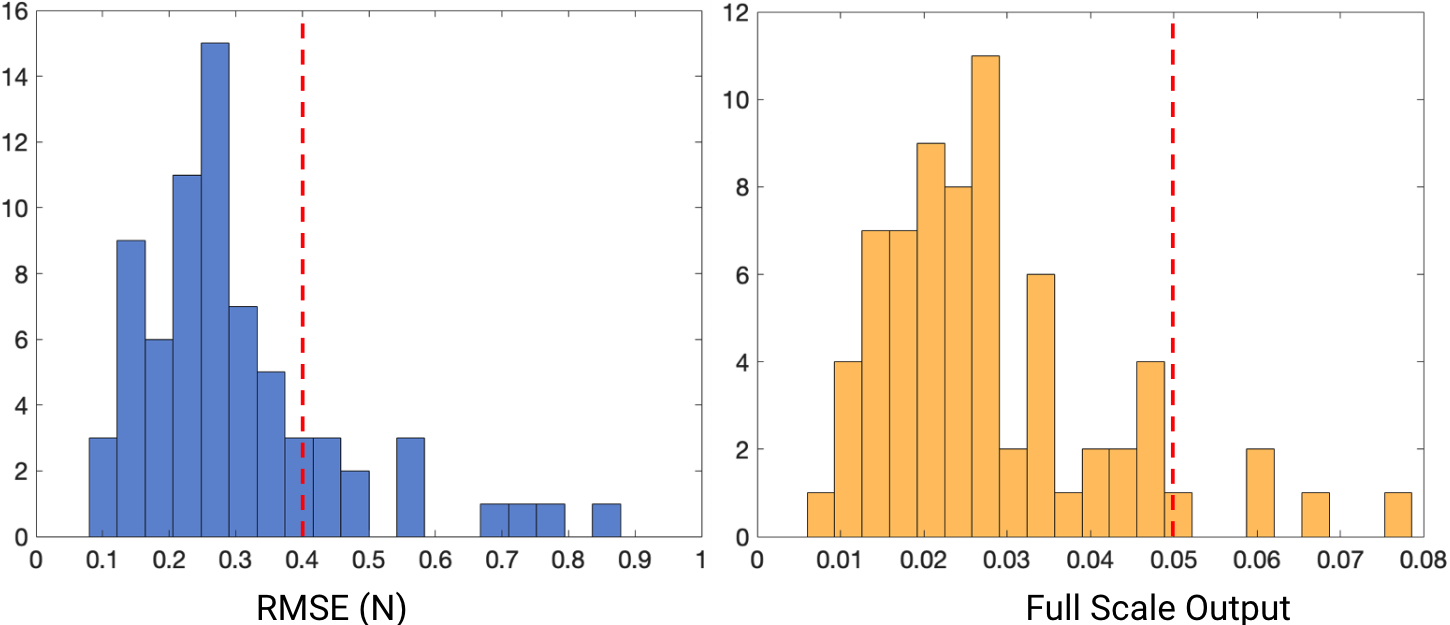}
\caption{ Histogram of average error in RMSE (left) and histogram of worst error for all modules (right). Red dotted lines show the desired 0.4 N average error in RMSE and 5\% worst error in FSO.}
\label{validation}
\end{figure}

\subsection{Force-to-sensor calibration}
We loaded and unloaded each of the 72 sensing modules between 0 and 25 N for 2 trials. Results for all modules exhibit a high linearity between sensor output and true force measured by the load cell (Fig. 7). Since stiffness varies with size of the flexures as fret width also varies, the calibration slopes of the five modules shows some differences.

\subsection{Validation}
To test the accuracy of the sensing modules, we apply random level of forces to the sensors using the micron stage and calculate the errors based on individual calibration curves. Fig. 8 shows the histograms for the average and worst error for all 72 sensing modules. We find that 81\% of the modules show an average error of \textless0.4 N RMSE, and 90\% of the modules show a worst error of \textless5\% FSO. The results satisfy the high accuracy desired for our purposes.

We believe that variances among sensing modules may be due to local temperature differences. To combat this issue, a temperature compensation system is installed within the current prototype: beyond the 6 photointerrupters at each fret, an extra one is placed in the middle and measures against a fixed surface. Assuming that temperature affects all photointerrupters in a fret the same way, the differential between each of the 6 photointerrupters with the fixed one should remain constant independent of the temperature. This compensation feature was tested independently but was not used during the user tests.

% ==================
% # Conclusion #
% ==================
\vspace{2mm}
\input\section{Conclusion and Future Work}
%\vspace{3mm}
This paper presents a novel compact, multi-contact force-sensing fretboard for a classical guitar. The 4mm thick design addresses the previous prototype's inability to distinguish individual forces on different fingers and the changes in the guitar's shape and weight. It was evaluated using a custom-designed test rig to demonstrate the desired accuracy and linearity. The prototype weighs and looks virtually identical to a regular fretboard once mounted. The functionality and feel of our system have also been evaluated by musicians from the Peabody Institute of the Johns Hopkins University, using a Likert scale questionnaire with a section for general comments. Users found the force sensing capabilities and real-time visual feedback constructive to their playing, despite noticing slight differences in sound quality. 

In the future more data will be collected and analyzed through user testing. Other directions include testing and evaluating the temperature compensation system, improving the update rate of the system and reducing electronic noise. 

%The entire system is contained in a 4 mm thick fretboard that can be mounted on a guitar using screws. Electro-mechanical design of the flexures and photointerruptor array relates force to a measurable voltage signal. Calibration and validation results show high linearity and accuracy. We also performed user testing with novice and expert musicians at the Peabody Institute, Johns Hopkins University (Fig. 8). %

%We find that signals of photointerrupters are susceptible to ambient temperature change. Our final design features temperature compensation using an extra sensing module for each fret by measuring to a fixed surface instead of to the flexure. However, we have not utilized the feature in the fretboard output.%

% Another issue we found during user testing is that the sound of the guitar is not as bright as that of a normal classical guitar. We suspect that it is due to the deflection of the flexure and the empty space between the flexure and the photointerrupter. This issue can be addressed by installing viscoelastic element below the flexure.

% ==================
% # Acknowledgment #
% ==================
\vspace{2mm}
\input\section*{Acknowledgment}
%\vspace{3mm}
We thank Melissa Hullman and Dr. Jing Xu for their valuable insights, luthier Garrett Lee for his help on building the final guitar, and Drs. Jae Kun Shim and Hyun Joon Kwon (Department of Kinesiology, University of Maryland) for earlier prototype development, preliminary data collection and analysis. The project was funded by private philanthropy through the Peabody Conservatory.

% ==============
% # REFERENCES #
% ==============
\bibliographystyle{IEEEtran}
\bibliography{biblio_rectifier}

\end{document}